\documentclass[preprint]{aastex}

\usepackage{graphicx,amssymb,amsmath,multirow,gensymb,lscape}
\usepackage{slashbox,enumitem,arydshln,multirow}
\usepackage{natbib}
\newcommand{\cden}{\mbox{cm$^{-2}$}} 
\newcommand{\um}{\mbox{$\mu$m}}
\newcommand{\Msun}{\mbox{M$_{\odot}$}}
\newcommand{\cmg}{\mbox{cm$^2$ g$^{-1}$}}
\newcommand{\Av}{\mbox{$A_V$}}
\newcommand{\NHH}{\mbox{N(H$_2$)}}

\newcommand{\Tbol}{\mbox{$T_{\rm bol}$}}
\newcommand{\Lbol}{\mbox{$L_{\rm bol}$}}
\newcommand{\Lsmm}{\mbox{$L_{\rm smm}$}}
\newcommand{\Lsun}{\mbox{$L_{\odot}$}}
\newcommand{\Menv}{\mbox{$M_{\rm env}$}}
\newcommand{\Nclass}{\mbox{$N_{Class\emph{0}}$}}

\shorttitle{Class 0 Protostars and Dense Gas in Perseus}
\shortauthors{Sadavoy et al.}

\title{Class 0 Protostars in the Perseus Molecular Cloud:  A Correlation Between the Youngest Protostars and the Dense Gas Distribution}
\author{S. I. Sadavoy\altaffilmark{1,2,3},
	J. Di Francesco\altaffilmark{1,2},
	Ph. Andr\'{e}\altaffilmark{4},
	S. Pezzuto\altaffilmark{5},
	J.-P. Bernard\altaffilmark{6,7},
	A. Maury\altaffilmark{4},
	A. Men'shchikov\altaffilmark{4},
	F. Motte\altaffilmark{4},
	Q. Nguy$\tilde{\hat{\rm e}}$n-Lu{\hskip-0.65mm\small'{}\hskip-0.5mm}o{\hskip-0.65mm\small'{}\hskip-0.5mm}ng\altaffilmark{8},
	N. Schneider\altaffilmark{9},
	D. Arzoumanian\altaffilmark{10}, 
	M. Benedettini\altaffilmark{5},
	S. Bontemps\altaffilmark{9,11},
	D. Elia\altaffilmark{5},
	M. Hennemann\altaffilmark{4},
	T. Hill\altaffilmark{12},
	V. K\"{o}nyves\altaffilmark{4,10},
	F. Louvet\altaffilmark{4},
	N. Peretto\altaffilmark{13},
	A. Roy\altaffilmark{4},
	G. J. White\altaffilmark{14,15}
	}
\altaffiltext{1}{Department of Physics \& Astronomy, University of Victoria, PO Box 355, STN CSC, Victoria, BC, V8W 3P6, Canada}
\altaffiltext{2}{National Research Council Canada, 5071 West Saanich Road, Victoria, BC, V9E 2E7, Canada}
\altaffiltext{3}{Current address: Max-Planck-Institut f\"{u}r Astronomie (MPIA), K\"{o}nigstuhl 17, 69117 Heidelberg, Germany; sadavoy@mpia.de}
\altaffiltext{4}{Laboratoire AIM, CEA/DSM-CNRS-Universit\'{e} Paris Diderot, IRFU/Service dÕAstrophysique, Saclay, 91191 Gif-sur-Yvette, France}
\altaffiltext{5}{Istituto di Astrofisica e Planetologia Spaziali, via Fosso del Cavaliere 100, 00133, Rome, Italy}
\altaffiltext{6}{CNRS, IRAP, 9 Av. colonel Roche, BP 44346, 31028 Toulouse Cedex 4, France}
\altaffiltext{7}{Universit\'{e} de Toulouse, UPS-OMP, IRAP, 31028 Toulouse Cedex 4, France}
\altaffiltext{8}{Canadian Institute for Theoretical Astrophysics, University of Toronto, Toronto, ON, M5S 3H8, Canada}
\altaffiltext{9}{Universit\'{e} de Bordeaux, LAB, UMR 5804, F-33270, Floirac, France}
\altaffiltext{10}{IAS, CNRS (UMR 8617), Universit\'{e} Paris-Sud 11, B\^{a}timent 121, 91400 Orsay, France}
\altaffiltext{11}{CNRS, LAB, UMR 5804, F-33270, Floirac, France}
\altaffiltext{12}{Joint ALMA Observatory, Alonso de Cordova 3107, Vitacura 763-0355, Santiago, Chile}
\altaffiltext{13}{School of Physics and Astronomy, Cardiff University, Cardiff, CF24 3AA, UK}
\altaffiltext{14}{Department of Physics and Astronomy, The Open University, Milton Keynes, MK7 6AA, UK}
\altaffiltext{15}{The Rutherford Appleton Laboratory, Chilton, Didcot, Oxfordshire, OX11 0QX, UK}

\begin{document}

\begin{abstract}
 
We use PACS and SPIRE continuum data at 160 \um, 250 \um, 350 \um, and 500 \um\ from the \emph{Herschel} Gould Belt Survey to sample seven clumps in Perseus: B1, B1-E, B5, IC348, L1448, L1455, and NGC1333.  Additionally, we identify and characterize the embedded Class 0 protostars using detections of compact \emph{Herschel} sources at 70 \um\ as well as archival \emph{Spitzer} catalogues and SCUBA 850 \um\ photometric data.  We identify 28 candidate Class 0 protostars, four of which are newly discovered sources not identified with \emph{Spitzer}.   We find that the star formation efficiency of clumps, as traced by Class 0 protostars, correlates strongly with the flatness of their respective column density distributions at high values.  This correlation suggests that the fraction of high column density material in a clump reflects only its youngest protostellar population rather than its entire source population.  We propose that feedback from either the formation or evolution of protostars changes the local density structure of clumps.

\end{abstract}


\section{Introduction\label{perseusIntro}}
  
Star formation appears connected to dense material within molecular clouds.  For example, dense cores are found towards the highest extinction material in their parent clouds (e.g., \citealt{Johnstone04}; \citealt{Kirk06}) and gravitationally-bound prestellar cores are primarily detected towards supercritical filaments (\citealt{Andre10}).  Similarly, \citet{Lada10} found a tight correlation between the number of young stellar objects (YSOs) detected in a cloud and the quantity of high extinction material towards that cloud (see also, \citealt{Heiderman10}).   

In a similar approach, \citet{Kainulainen09} demonstrated that quiescent molecular clouds have extinction probability density functions (PDFs) with log-normal shapes whereas molecular clouds actively forming stars have additional power-law extensions (tails) at high values.  This distinction between quiescent clouds and active clouds has been seen in a variety of systems and studies (e.g., \citealt{Kainulainen11}; \citealt{Hill11}; \citealt{Schneider12}).  The origin of these power-law tails is still unknown.  \citet{Kainulainen11} determined that the material associated with them were pressure confined, whereas \citet{Schneider13} argued these tails are due to gravity.  To understand better how these power-law tails correlate with star formation, we  compared them to youngest generations of YSOs (Class 0) in clouds.  

Class 0 protostars are deeply embedded in thick envelopes and are mainly characterized by significant emission at long wavelengths (\citealt{Andre93}).  Due to these envelopes, Class 0 protostars are difficult to classify and several techniques have been used to distinguish Class 0 protostars from later-stage Class I sources, including low bolometric temperatures, \Tbol\ (e.g., \citealt{Chen95}; \citealt{Evans09}), high ratios of submillimeter luminosity to bolometric luminosity, \Lsmm/\Lbol\ (e.g., \citealt{Andre00}; \citealt{Stutz13}), or envelope masses, \Menv, that dominate over the stellar masses, $M_{\star}$ (e.g.,\citealt{AndreMontmerle94}; \citealt{Bontemps96}; \citealt{Maury11}).  

 In this Letter, we combine \emph{Herschel} (\citealt{Pilbratt10}) continuum observations from the \emph{Herschel} Gould Belt Survey (HGBS; \citealt{Andre10}) with archival \emph{Spitzer} YSO catalogues and 850 \um\ SCUBA data for the Perseus molecular cloud.  Perseus is relatively nearby ($\sim$ 235 pc; \citealt{Hirota08}) and contains rich populations of dense cores and YSOs  indicative of both low- and intermediate-mass star formation within clustered and isolated environments (\citealt{Bally08}). In addition, Perseus has seven clumps (B1, B1-E, B5, IC348, L1448, L1455, and NGC1333) that provide a variety of environments and star formation histories, allowing us to compare and contrast Class 0 sub-populations within the same cloud.

\section{Data}\label{perseusObs}

\subsection{Herschel Observations}

We present the HGBS observations of Eastern Perseus, which cover $\sim 7$ deg$^2$ roughly centered on IC348 and were completed in 2011 February in the same manner as the observations of Western Perseus (see \citealt{Sadavoy12}; \citealt{Pezzuto12}).   These data utilized parallel observations with the PACS (\citealt{Poglitsch10}) and SPIRE (\citealt{Griffin10}) instruments, resulting in simultaneous mapping at 70 \um, 160 \um, 250 \um, 350 \um, and 500 \um\ with a 60 arcsec s$^{-1}$ scan rate.  In brief, the PACS and SPIRE data were reduced with modified scripts by M. Sauvage (PACS) and P. Panuzzo (SPIRE) using HIPE 7 and the PACS Calibration Set v26 and the SPIRE Calibration Tree 6.1, respectively.  The final maps were produced using \emph{scanamorphos} 11 (\citealt{Roussel12}).

\subsection{Archival Data}\label{archival}

We used archival submillimeter continuum observations from the SCUBA Legacy Catalogue (\citealt{difran08}) and archival YSO catalogues from \citet[][c2d catalogue]{Evans09} and \citet[][G09 catalogue]{Gutermuth09}.  For the SCUBA data, we used the publicly available, smoothed 850 \um\ maps at $\sim 23$\arcsec\ resolution.  For the YSO catalogues, we used the combined YSO sample in the G09 catalogue (293 YSOs for IC348 and NGC1333 only) and all sources in the c2d catalogue with either a ``YSOc'' or a ``red'' designation with an additional ``rising'' source that is a known YSO (651 YSOs), where 205 sources were common to both catalogues.   Both catalogues used data at 1.25 - 24 \um\ (2MASS and \emph{Spitzer}), but they employed different techniques to identify YSOs and remove contaminants.  As a result, sources could be identified as YSOs in one and not the other (\citealt{Gutermuth09}).


\section{Results}\label{perseusResults}

\subsection{Clumps}\label{clumps}

We fitted determined the line-of-sight averaged temperatures and column densities across the entire Perseus cloud in a pixel-by-pixel manner.  We corrected the zero-point offset in the Western and Eastern fields separately using  Planck data (e.g., \citealt{Bernard10}) and convolved all maps to a common resolution of 36.3\arcsec\ with 14\arcsec\ pixels.  The spectral energy distribution (SED) over $160-500$ \um\ of each pixel was fit with a modified black body function of $I_{\nu} = \kappa_{\nu}B(\nu,T)\Sigma$, where $\kappa_{\nu} \sim \nu^{\beta}$ is the dust opacity, $B(\nu,T)$ is the black body function at temperature $T$, and $\Sigma = \mu m_p \NHH$\ is the gas mass column density for a mean molecular weight per hydrogen molecule of $\mu = 2.8$ (e.g., \citealt{Kauffmann08}).  We assumed a fixed dust opacity law with a dust emissivity index $\beta = 2$ and $\kappa = 0.1\ \cmg$\ at 300 \um.  We also applied average colour correction factors assuming $\beta = 2$ and temperatures between 10 K and 25 K.  

\begin{figure}[h!]
\includegraphics[scale=1]{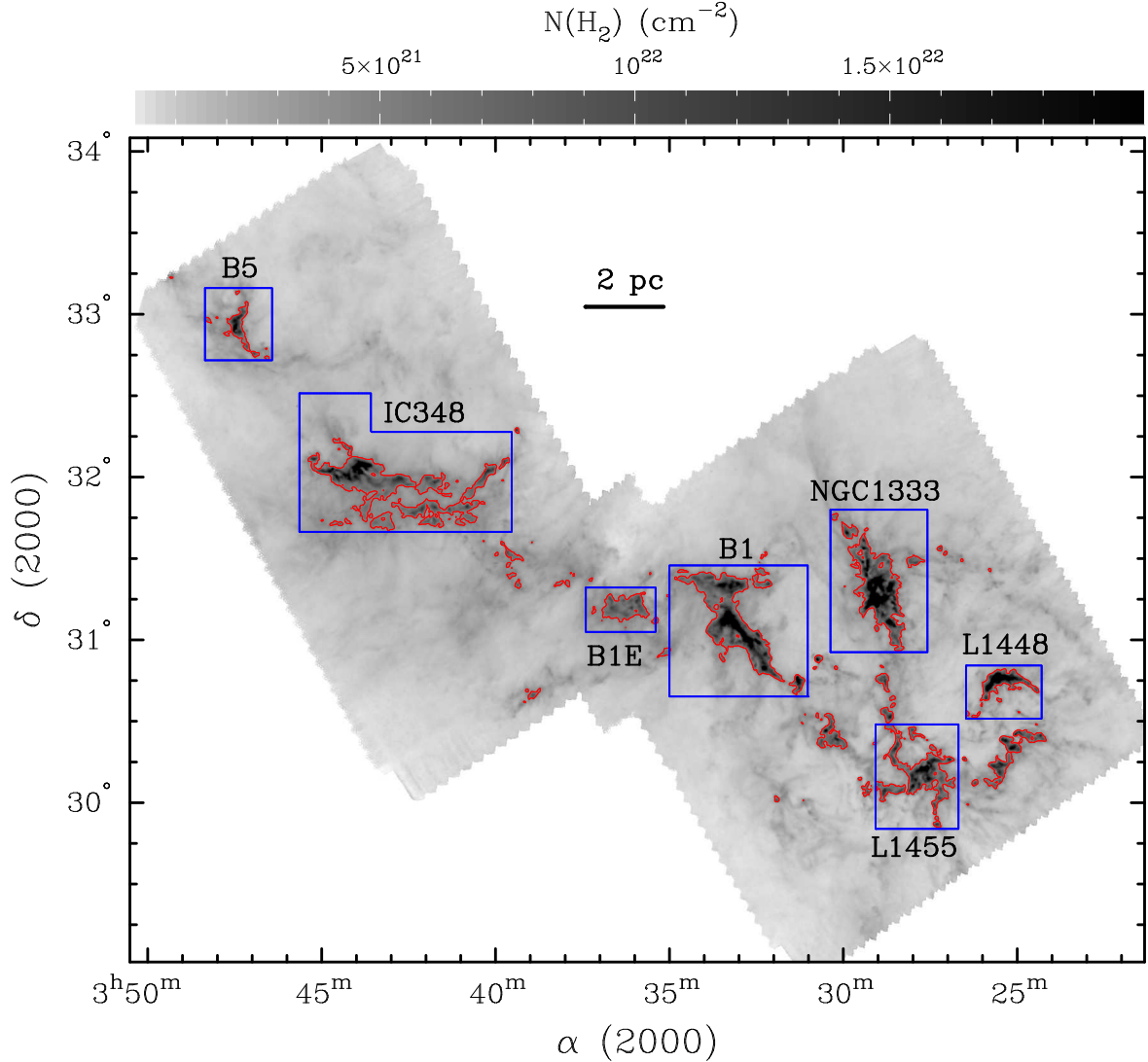}
\caption{Column density map derived from HGBS data of the Perseus molecular cloud.  Contours correspond to $\NHH = 5 \times 10^{21}$ \cden. The boxes illustrate the boundaries selected for each clump. }\label{clump_map}
\end{figure}

Figure \ref{clump_map} shows the \emph{Herschel}-derived column density map for Perseus with the boundaries adopted for each clump.  We defined these boundaries to enclose high column density material towards each clump, taking $A_V = 7$ as our threshold (see \citealt{Lada10}; \citealt{Hill11}).  An $A_V = 7$ mag translates to $\NHH \sim 5-7 \times 10^{21}$ \cden\ using either the conversion factors of $\NHH/A_V = 9.4 \times 10^{20}\ \cden\ \mbox{mag}^{-1}$ (\citealt{Bohlin78}) or $\NHH/A_V = 6.9 \times 10^{20}\ \cden\ \mbox{mag}^{-1}$ (\citealt{Draine03}; \citealt{Evans09}), where the former factor corresponds better with lower column densities (e.g., $\lesssim 6 \times 10^{21}$ \cden) and the latter with higher column densities (e.g., $\gtrsim 6 \times 10^{21}$ \cden; see also, \citealt{Roy14}).  Since $\Av \approx 7$ falls near the transition between these two conversion factors, we adopted the lower value of $\NHH = 5 \times 10^{21}$ \cden\ to define the boundaries.  For IC348, we modified slightly the box boundary to include all sources in the G09 catalogue.  Everything outside these boundaries are considered ``Off Clump.'' 

\begin{figure}[h!]
\begin{tabular}{ll}
\includegraphics[scale=0.85,trim=1mm 2mm 2mm 2mm,clip=true]{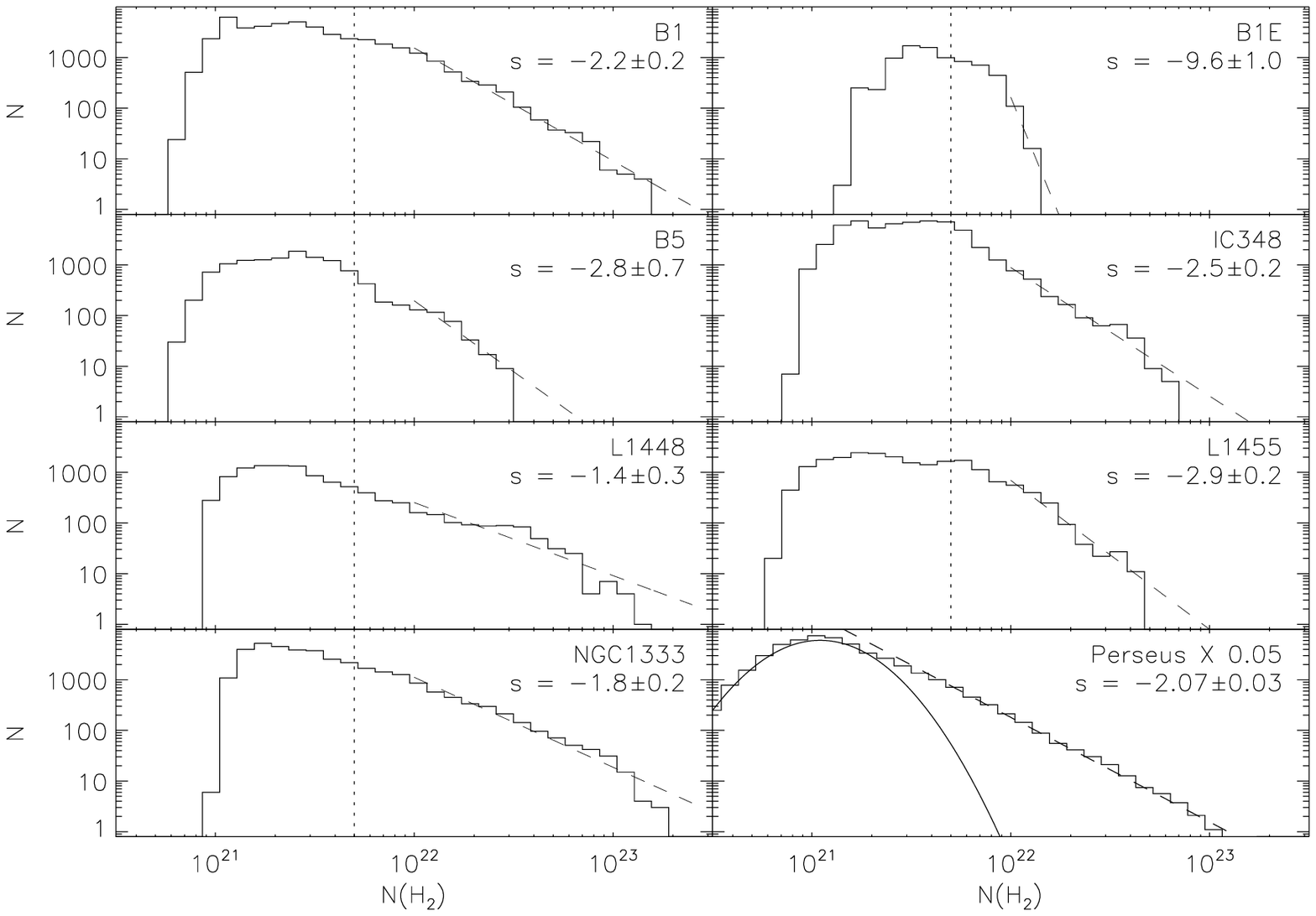}
\end{tabular}
\caption{Column density PDFs for each of the Perseus clumps and the entire Perseus cloud (scaled down by a factor of 0.05).  For the clumps, the PDFs include all material within their respective boundaries shown in Figure \ref{clump_map}, and are incomplete for $\NHH \lesssim 5 \times 10^{21}$ \cden\ (shown by the dotted lines).  The dashed lines give the linear least squares best-fit slope, $s$, for $\NHH \ge 1 \times 10^{22}$ \cden.  For the entire Perseus cloud PDF, we include a best-fit log-normal distribution and power-law tail. }\label{pdfs}
\end{figure}

Figure \ref{pdfs} shows the column density PDFs of each clump for all material within their respective boundaries and for the entire Perseus molecular cloud.  We fitted the PDF for the entire Perseus cloud with a log-normal distribution and a single power-law tail (e.g., as in \citealt{Kainulainen09}; \citealt{Schneider12}).  The clump PDFs, however, cannot be fit by a log-normal as they are incomplete at low values (i.e., $\NHH \lesssim 5 \times 10^{21}$ \cden).  Nevertheless, we note that some clumps have significant tails at high values, whereas others have little to no extensions.  For each clump, we measured the power-law slopes of the PDF tails using linear least squares fits for $\NHH\ \ge 1.0 \times\ 10^{22}\ \cden$.  This threshold represents well the power-law tail for the entire Perseus cloud and should be well-sampled towards the clumps. Using different bin sizes, we found that the power-law slopes have uncertainties of $\lesssim 20$\%.  

\subsection{Class 0 Protostars}\label{herschelSources}

\subsubsection{Source Extraction}\label{getsources}

We identified sources in the \emph{Herschel} $70-500$ \um\ maps using the \emph{getsources} algorithm (\citealt{getsources}).  We also used \emph{getsources} to obtain follow-up 850 \um\ fluxes from the SCUBA maps.  Since SCUBA only mapped $\sim 3.5$ deg$^2$ of Perseus (\citealt{Kirk06}) whereas \emph{Herschel} mapped $\sim 15$ deg$^2$, we cannot obtain 850 \um\ data for all \emph{Herschel} sources.  Therefore, we only measured 850 \um\ fluxes for sources with low bolometric temperatures based on their \emph{Spitzer} and \emph{Herschel} data alone (see Section \ref{source_class}).  For these sources, 850 \um\ fluxes were measured at the position of each candidate using our \emph{getsources}-derived source footprints.

\subsubsection{Class 0 Identification}\label{source_class} %

A source was considered protostellar if it was associated with either (1) a compact 70 \um\ detection indicative of warm envelopes internally heated by YSOs (e.g., \citealt{Dunham08}; \citealt{Bontemps10}) or (2) a \emph{Spitzer}-identified YSO from the c2d or G09 catalogues within 8\arcsec\ (roughly the effective radius of the 160 \um\ beam).   Compact 70 \um\ sources were identified with sizes $\le 13\arcsec$\ (roughly 1.5 times the PACS 70 \um\ beam), low elongations ($a/b \le 2$), and signal-to-noise ratios $\ge 3$.  We visually inspected all sources in the original 70 \um\ maps, adding seven compact detections.   

These protostellar signatures include later-stage YSOs and non-YSO contaminants (e.g., see \citealt{Dunham08}).  To ensure a robust catalogue of Class 0 sources, we required that all candidates have (1) $\Tbol\ < 70$ K, (2) $\Lsmm/\Lbol\ \ge 1$\%, and (3) $\Menv > M_{\star}$ (e.g., \citealt{Chen95}; \citealt{Andre00}; \citealt{Evans09}; \citealt{Maury11}).  \Tbol\ and \Lbol\ were determined from trapezoidal integration across all observed bands over $1.25 - 850$ \um\ (e.g., \citealt{Dunham08}), whereas \Lsmm\ was determined from trapezoidal integration for $\lambda \ge 350$ \um, only.  We determined \Menv\ directly from fitting the getsources-determined fluxes at $160-850$ \um\ with,
\begin{equation}
S_{\nu} =  \kappa_{\nu} B(\nu,T) \Menv/D^2,    \label{sourceEq}
\end{equation}
\noindent where $D$ is the distance and $\kappa_{\nu}$ is the same as in Section \ref{clumps}.   Finally, we compared \Menv\ and \Lbol\ using the same method as \citet{Maury11} to identify those sources with significant envelope mass.  Briefly, we used accretion models to characterize the evolution of \Menv\  versus $M_{\star}$ (as determined by \Lbol) from Class 0 to Class I using an accretion efficiency of 50\%.  Based on these models, we identified Class 0 sources using $\Menv/\Msun > 0.2(\Lbol/\Lsun)^{0.6}$ and $\Menv/\Msun > 0.1 \Lbol/\Lsun$, where the former assumes a decreasing accretion rate (with \Menv) and the latter assumes a constant accretion rate (e.g., see \citealt{AndreMontmerle94}; \citealt{Bontemps96}; \citealt{Maury11}).

We identified 28 sources as Class 0 candidates.  Table \ref{app_class0} gives their locations (parent clump), \emph{Herschel} source names, coordinates, indications of corresponding \emph{Spitzer}-identified YSOs, and measurements for \Tbol, \Lbol, \Lsmm, \Lsmm/\Lbol, and \Menv.  Source names are taken from the observation field (West or East) and the \emph{getsources} detection number. The errors quoted in Table \ref{app_class0} were determined from applying 1000 random factors corresponding to the flux errors, calibration errors (assuming 20\%\ at 850 \um, 10\%\ elsewhere), and \emph{Herschel} colour uncertainties (i.e., \citealt{Sadavoy13}).  These errors reflect only the observational uncertainties, however, and should be considered lower limits of the absolute error.  In particular, a fixed dust opacity law can result in mass or column density uncertainties of factors of $1.5-2$ (e.g., \citealt{Roy13,Roy14}; \citealt{Ysard13}).

\begin{deluxetable}{llcccccccc}
\tabletypesize{\scriptsize}
\tablewidth{0pt}
\tablecolumns{8}
\tablecaption{Candidate Class 0 Protostars in Perseus\label{app_class0}}
\tablehead{
\colhead{Clump} &
\colhead{Source} &
\colhead{RA} &
\colhead{Dec} &
\colhead{Spitzer\tablenotemark{a}} &
\colhead{\Tbol\tablenotemark{b}} &
\colhead{\Lbol\tablenotemark{b}} & 
\colhead{\Lsmm\tablenotemark{b}} &
\colhead{\Lsmm/\Lbol\tablenotemark{b}} &
\colhead{$M_{env}$\tablenotemark{b}} \\
	&
	&
(J2000) &
(J2000) &
	&
(K)	&
(\Lsun) &
($10^{-2}$ \Lsun) &
(\%)	&
(\Msun)
}
\startdata
\hline
L1448      & West9    & 3:25:22.3    &	30:45:10	& Y   					& $43 \pm 2$ 	& $3.6 \pm 0.5$ 	& $9.5 \pm 0.9$  & $2.7 \pm 0.4$	&  $0.8 \pm 0.2$   \\
L1448      & West25  & 3:25:35.4    & 30:45:32	& N                   		  	& $22 \pm 1$ 	& $1.4 \pm 0.1$ 	& $11  \pm 1$ 	  & $7.8 \pm 0.9$  	&  $1.5 \pm 0.3$    \\
L1448      & West8    & 3:25:36.2    & 30:45:17	& Y   					& $57 \pm 3$ 	& $8.3 \pm 0.8$ 	& $28  \pm 2$ 	  & $3.4 \pm 0.4$	&  $3.1 \pm 0.6$    \\
L1448      & West4    & 3:25:38.7    & 30:44:02	& Y   					& $47 \pm 2$ 	& $9.2 \pm 1.3$ 	& $18  \pm 2$ 	  & $1.9 \pm 0.3$	&  $1.3 \pm 0.2$   \\
L1455      & West18  & 3:27:43.1    & 30:12:26	& Y			 		& $65 \pm 3$ 	& $1.4 \pm 0.2$ 	& $5.0 \pm 0.5$  & $3.6 \pm 0.5$	&  $0.5 \pm 0.1$    \\
NGC1333    & West162  & 3:28:38.6  & 31:06:00	& Y   				& $22 \pm 1$ 	& $0.3 \pm 0.02$	& $6.1 \pm 0.6$  & $21 \pm 3$		&  $1.8 \pm 0.3$   \\
NGC1333    & West33   & 3:29:00.4    & 31:11:57	& Y				   & $32 \pm 2$ 	& $0.5 \pm 0.06$ 	& $3.1 \pm 0.7$  & $6 \pm 2$		&  $0.4 \pm 0.1$   \\
NGC1333    & West19   & 3:29:01.8    & 31:15:34	& N	                    	   & $21 \pm 1$ 	& $1.5 \pm 0.2$ 	& $18  \pm 2$ 	  & $12 \pm 2$		&  $3.1 \pm 0.5$    \\
NGC1333    & West40   & 3:29:03.9    & 31:14:43	& Y				   & $36 \pm 4$ 	& $0.4 \pm 0.1$ 	& $6.6 \pm 0.9$  & $18 \pm 5$		&  $1.6 \pm 0.4$    \\
NGC1333    & West87\tablenotemark{c}	   & 3:29:06.7    & 31:15:33 & Y    & $\lesssim 23$ & $\lesssim 0.3$  	& $7.8 \pm 1.0$  & $\gtrsim 26$	&  $1.9 \pm 0.5$    \\
NGC1333    & West6    & 3:29:10.3     & 31:13:28	& Y			             & $29 \pm 2$ 	& $7.0 \pm 0.7$ 	& $35 \pm 3$   	   & $5.1 \pm 0.7$	&  $8 \pm 2$    \\
NGC1333    & West14   & 3:29:11.0    & 31:18:26	& Y				   & $46 \pm 3$ 	& $3.6 \pm 0.5$ 	& $11  \pm 1$   	  & $2.9 \pm 0.5$	&  $1.1 \pm 0.2$   \\
NGC1333    & West13   & 3:29:11.9    & 31:13:05	& Y				   & $28 \pm 1$ 	& $4.0 \pm 0.3$ 	& $22 \pm 2$      & $5.5 \pm 0.7$	&  $3.0 \pm 0.5$    \\
NGC1333    & West30   & 3:29:13.5    & 31:13:54	& Y				   & $31 \pm 2$ 	& $0.7 \pm 0.08$ 	& $3.7 \pm 1.5$  & $6 \pm 2$		&  $0.3 \pm 0.1$   \\
NGC1333    & West23   & 3:29:17.2    & 31:27:43	& Y				   & $42 \pm 2$ 	& $0.8 \pm 0.1$ 	& $2.8 \pm 0.4$  & $3.5 \pm 0.6$	&  $0.3 \pm 0.1$    \\
NGC1333    & West37   & 3:29:18.8    & 31:23:12	& N	                     	    & $22 \pm 1$ 	& $0.5 \pm 0.1$ 	& $3.4 \pm 0.7$  & $7 \pm 3$		&  $0.5 \pm 0.1$    \\
NGC1333    & West28   & 3:29:51.7    & 31:39:03	& Y				   & $36 \pm 2$ 	& $0.6 \pm 0.06$ 	& $3.5 \pm 0.3$  & $6.0 \pm 0.8$	&  $0.8 \pm 0.2$    \\
B1         & West17   & 3:31:20.6    &	30:45:29	& Y					   & $32 \pm 2$ 	& $1.3 \pm 0.1$ 	& $6.6 \pm 0.6$  & $5.1 \pm 0.7$	&  $0.8 \pm 0.2$    \\
B1         & West26   & 3:32:17.7    &	30:49:46	& Y					   & $27 \pm 1$ 	& $0.9 \pm 0.07$ 	& $7.9 \pm 0.7$  & $9 \pm 1$		&  $2.4 \pm 0.6$    \\
B1         & West50   & 3:33:14.3    &	31:07:09	& Y					   & $52 \pm 3$ 	& $0.3 \pm 0.03$ 	& $3.6 \pm 0.6$  & $13 \pm 3$		&  $0.7 \pm 0.2$    \\
B1         & West34   & 3:33:16.2    &	31:06:51	& Y					   & $30 \pm 2$ 	& $0.6 \pm 0.05$ 	& $5.7 \pm 0.7$  & $10 \pm 1$		&  $0.7 \pm 0.2$    \\
B1         & West12   & 3:33:17.7    & 31:09:30	&  Y					   & $48 \pm 1$ 	& $3.7 \pm 0.4$ 	& $14  \pm 1$      & $3.6 \pm 0.5$	&  $1.7 \pm 0.3$    \\
B1	     & West41\tablenotemark{d}	  &  3:33:21.3 & 31:07:27	& N	            & $19 \pm 1$ 	& $0.7 \pm 0.07$ 	& $13  \pm 1$      & $18 \pm 3$		&  $3.3 \pm 0.5$  \\
IC348      & East11   & 3:43:50.6    &	32:03:24	& Y				   	& $39 \pm 2$ 	& $0.4 \pm 0.04$ 	& $3.4 \pm 0.5$   & $9 \pm 2$		&  $0.5 \pm 0.1$    \\
IC348      & East9    & 3:43:50.6    &	32:03:08	& Y				   	& $45 \pm 2$ 	& $0.7 \pm 0.08$ 	& $3.0 \pm 0.6$   & $5 \pm 1$		&  $0.3 \pm 0.1$    \\
IC348      & East4    & 3:43:56.4    &	32:00:49	& Y				   	& $27 \pm 1$ 	& $1.8 \pm 0.1$ 	& $12 \pm 1$       & $6.7 \pm 0.8$	&  $2.4 \pm 0.5$    \\
IC348      & East5    & 3:43:56.6    &	32:03:03	& Y				   	& $30 \pm 2$ 	& $1.5 \pm 0.1$ 	& $9.4 \pm 0.09$ & $6.4 \pm 0.9$	&  $1.1 \pm 0.2$    \\
IC348      & East17   & 3:44:02.1   &  32:02:02	& Y				   	& $57 \pm 10$ 	& $0.3 \pm 0.1$ 	& $2.5 \pm 0.04$ & $8 \pm 3$		&  $0.8 \pm 0.3$    \\
\enddata
\tablenotetext{a}{Indicates a \emph{Spitzer}-identified YSO in either the c2d (\citealt{Evans09}) or G09 (\citealt{Gutermuth09}) catalogues.}
\tablenotetext{b}{Observed source properties (see text for details).   Errors correspond to 1 $\sigma$ uncertainties based on 1000 random corrections within the observational uncertainties only.  These errors are lower-limits.} 
\tablenotetext{c}{West87 lacked a compact 70 \um\ source, possibly due to diffuse 70 \um\ emission.  We used a 1 $\sigma$\ point source upper limit of 100 mJy at 70 \um\ to calculate \Lbol\ and \Tbol. (Note that the associated \emph{Spitzer} source has a 24 \um\ flux of only $\sim 3$ mJy.)}
\tablenotetext{d}{A compact 70 \um\ source was detected with follow-up PACS 70 \um\ data at a 20 arcsec s$^{-1}$ scan rate (Pezzuto et al., in preparation).}
\end{deluxetable}

Table \ref{app_class0} includes only well-characterized Class 0 protostars.  For example, we excluded three Class 0-like sources that were not observed by SCUBA since their values for \Lsmm\ and \Menv\ are less constrained.   Additionally, we used $\Lsmm/\Lbol > 1\%$\ rather than 0.5 \%\ (e.g., \citealt{Andre93}) to exclude borderline objects and concentrate on robust Class 0 protostars.   If we relaxed our criteria (i.e., $\Lsmm/\Lbol \ge 0.5$\%; either $\Menv/\Msun > 0.2(\Lbol/\Lsun)^{0.6}$ or $\Menv/\Msun > 0.1\Lbol/\Lsun$), we would include only four additional sources. 

Four of our 28 Class 0 candidates have no \emph{Spitzer}-identified YSOs within 8\arcsec.   A YSO could be missed by \emph{Spitzer} if it is (1) displaced from its natal core, (2) misclassified in the literature, (3) very faint, or (4) obscured by nearby bright emission.   Of these four, West25, West19, and West37 are located towards halos of brighter YSOs or bright diffuse emission, and thus, a compact \emph{Spitzer} source may be lost.  Class 0 protostars are relatively fainter in the \emph{Spitzer} bands than later-stage YSOs and will be  intrinsically more difficult to detect towards diffuse emission or halos (e.g., \citealt{Rebull07}).  The fourth source, West41, corresponds to a first hydrostatic core candidate (\citealt{Pezzuto12}) and is unlikely to have a \emph{Spitzer} counterpart.   

Table \ref{clumpPopTable} gives the clump PDF power-law slope (from Figure \ref{pdfs}), mass, area, and the number of associated Class 0 objects  as determined only from material at $\NHH \ge 1 \times 10^{22}$ \cden.  Clump masses were measured from their total column densities with $M = \mu m_p \Omega D^2 \NHH$, where $\Omega$ is the solid angle of the pixels, and their areas were measured from the number of pixels with $\NHH \ge 1 \times 10^{22}$ \cden.  Table \ref{clumpPopTable} also includes the results for the Perseus cloud as a whole.  

\begin{table}[h!]
\caption{Perseus Clump Properties}\label{clumpPopTable}
\begin{tabular}{lcccc}
\hline\hline
Clump		& s\tablenotemark{a} 	&  Mass\tablenotemark{a} (\Msun)	& Area\tablenotemark{a}	(pc$^2$) & \Nclass	\\
\hline
L1448     		& $-1.4 \pm 0.3$		&	118		&	0.21	&    4  	\\			
NGC1333		& $-1.8 \pm 0.2$		&	365		&	0.73	&  12  	\\  			
B1          		& $-2.2 \pm 0.2$		&   	342		& 	0.84	&    6 	\\			
IC348     		& $-2.5 \pm 0.2$		&	156		&	0.44	&    5  	\\			
B5           		& $-2.8 \pm 0.7$		&	28		&	0.09	&    0   	\\			
L1455      		& $-2.9 \pm 0.2$		&	101		&	0.31	&    1  	\\  			
B1-E        		& $-9.6 \pm 1.0$		&   	5		&	0.02	&    0   	\\			
Perseus		& $-2.07 \pm 0.03$		& 	1171		& 	2.8	&  28		\\
\hline
\end{tabular}
\tablenotetext{a}{Corresponding to material with $\NHH \ge 1 \times 10^{22}$ \cden\ only, where the PDF tail slope, $s$ is given by $s = d\log{N}/d\log{\NHH}$ for $\NHH \ge 1 \times 10^{22}$ \cden\ (see Figure \ref{pdfs}).}
\end{table}

\section{Discussion}\label{persDiscussion}

\begin{figure}[h!]
\includegraphics[scale=0.8]{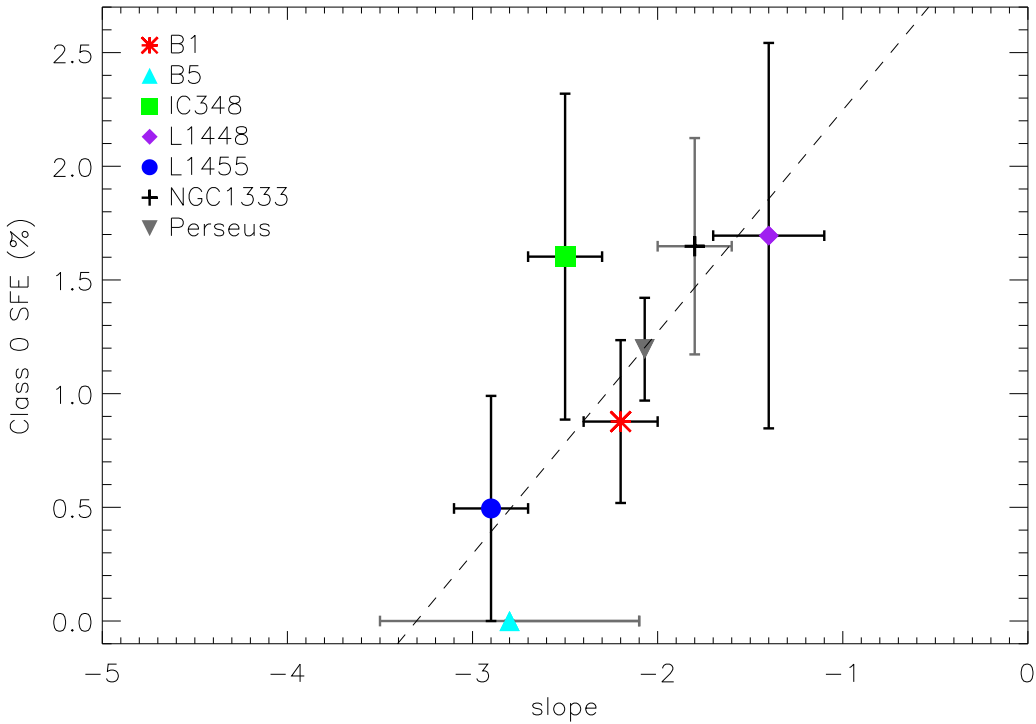}
\caption{Comparison of PDF power-law slopes from Figure \ref{pdfs} with Class 0 SFE, assuming a mean stellar mass of 0.5 \Msun.  The dashed line corresponds to a linear least squares best fit to the data (excluding B1-E).  We find a best-fit linear slope of $\sim 0.01$. The error bars indicate the robustness of the high column density slope (see Section \ref{clumps}) and Poisson uncertainties for the number of Class 0 sources ($\sqrt{N}$).}\label{clumpYoungFig}
\end{figure}

Figure \ref{clumpYoungFig} plots the slope of the PDF power-law tail for the Perseus clumps and the entire Perseus cloud against their respective Class 0 star formation efficiency (SFE), where SFE $= M_{YSO}/(M_{YSO} + M_{clump})$.  We adopted $M_{YSO} = \Nclass\ \times \langle M\rangle$, assuming a mean mass of $\langle M\rangle = 0.5\ \Msun$\ (e.g., \citealt{Evans09}).  In general, we find that clumps with flatter power-law tails are more efficient at forming Class 0 protostars (see also, \citealt{Bontemps10frag}), with a linear least squares best-fit of $\sim 0.01$ for all clumps excluding B1-E.  (B1-E is off the scale of Figure \ref{clumpYoungFig}.)   Extrapolating our linear relation in Figure \ref{clumpYoungFig}, power-law slopes of $s \gtrsim -3.5$ appear necessary to form Class 0 protostars.  Thus, clumps with very little high column density material (e.g., B1-E) need to undergo further compression to form stars.  

Figure \ref{clumpYoungFig} demonstrates that the PDF power-law slope relates to the youngest YSO populations, which agrees well with a recent study by Louvet et al. (2014, submitted), where they found a strong relationship between the dense core formation efficiency and the local volume density in W43 (see also \citealt{FederrathKlessen12}).  We note that the correlation in Figure \ref{clumpYoungFig} is unchanged if the column densities associated with the protostars in each clump are removed, indicating we are not biased by them.  Moreover, we retain the same correlation if the borderline Class 0 sources and non-SCUBA candidates (see Section \ref{source_class}) are included.   We also find a similar correlation ($s \sim 0.02$) with $M_{YSO} = \Sigma(0.5 \times \Menv)$, assuming an accretion efficiency of 50\%.  Finally, while previous studies have shown a close relation between instances of YSOs (all stages) and dense material (e.g., \citealt{Lada10}; \citealt{Heiderman10}),  we find no correlations between the power-law slope and the SFEs for later-stage YSOs (i.e., Class I, Class II), as defined by infrared spectral indices following \citet{Evans09}.  (Final catalogues for all sources will be presented in Pezutto et al., in preparation.)  These comparisons suggest that the correlation in Figure \ref{clumpYoungFig} is relatively robust and unique to Class 0 protostars. Therefore, \emph{we propose that the PDF power-law tail of a clump reflects only its youngest protostars} rather than its entire YSO population, where clumps with shallow high-value slopes (e.g., L1448 and NGC1333) are most actively forming stars at the present epoch.  

Several recent simulations have explored the formation of high density structures in clouds.  For example, simulations of gravitationally dominated clouds naturally produce flat, power-law tails at high densities (e.g., \citealt{BallesterosParedes11};  \citealt{Kritsuk11}; \citealt{Girichidis13}).  Alternatively, recent studies of high-mass star-forming regions have attributed high SFEs and densities to compression from colliding flows or expanding HII regions (e.g., \citealt{NguyenLuong11}; \citealt{Minier13}).  Nevertheless, most simulations generally exclude feedback, which may still have a significant impact (\citealt{HeitschHartmann08}).  For example, stellar feedback or YSO outflows may affect cloud material on clump scales (e.g., \citealt{Arce10}; \citealt{Colin13}).   Such feedback may explain why the power-law tail correlates only with Class 0 protostars.

Including feedback, star-forming regions may evolve as follows:  First, clumps form with initially log-normal column density distributions and as gravity dominates, they produce steep high (column) density tails.  Star formation in the clumps then proceeds either after or concurrently with the build-up of sufficient high column density material.  Next, feedback from the evolving YSO populations affects the clumps, e.g., by increasing the turbulence of the clump or ejecting dense material.  This feedback may effectively lower the fraction of high (column) density material in a clump, subsequently producing fewer Class 0 protostars (see Figure \ref{clumpYoungFig}).  Alternatively, if YSO feedback results in density enhancements (e.g., \citealt{Schneider13}), then clumps with more Class 0 protostars are more compressed, resulting in shallower high column density slopes.  In this scenario,  Class 0 protostars shape their local (column) density environment rather than the local environment regulating the Class 0 SFE.  Thus, the differences between the clumps in Figure \ref{clumpYoungFig} can be explained by YSO feedback, where Class 0 protostars either locally disrupt or enhance dense material.  
 
This study demonstrates the tight connection between high column density material and the youngest protostellar sources for a single molecular cloud.  Nevertheless, several of the clumps in Perseus have small Class 0 object samples.  Also, Perseus is not forming high-mass stars, which may yield a different relationships between the fraction of Class 0 protostars and the PDF slope.  With the full HGBS, we can explore further the relationship between source populations and clumps in a large variety of clouds and environments and test the effects of feedback.  These observations will improve constraints for simulations and models of molecular clouds that are necessary to reveal the evolution of dense structures and their relation to Class 0 protostars.

\vspace{1cm}
\acknowledgments{\noindent \emph{Acknowledgements:} This work was possible with funding from the Natural Sciences and Engineering Research Council (NSERC) Canadian Graduate Student award.   We acknowledge the support by the Canadian Space Agency (CSA) via a Space Science Enhancement Program grant, the NSERC via a Discovery grant, the National Research Council of Canada (NRC), and the European Research Council under the European UnionÕs Seventh Framework Programme (FP7/2007-2013 Ð ERC Grant Agreement no. 291294).   We thank the anonymous referee for improving the clarity of this paper.   SIS thanks M. Dunham, J.  Kainulainen, and A. Stutz for very useful discussions. 	\emph{Herschel} is an ESA space observatory with science instruments provided by European-led Principal Investigator consortia and with important participation from NASA.       PACS has been developed by a consortium of institutes led by MPE (Germany) and including UVIE (Austria); KU Leuven, CSL, IMEC (Belgium); CEA, LAM (France); MPIA (Germany); INAF-IFSI/OAA/OAP/OAT, LENS, SISSA (Italy); IAC (Spain). This development has been supported by the funding agencies BMVIT (Austria), ESA-PRODEX (Belgium), CEA/CNES (France), DLR (Germany), ASI/INAF (Italy), and CICYT/MCYT (Spain).        SPIRE has been developed by a consortium of institutes led by Cardiff University (UK) and including Univ. Lethbridge (Canada); NAOC (China); CEA, LAM (France); IFSI, Univ. Padua (Italy); IAC (Spain); Stockholm Observatory (Sweden); Imperial College London, RAL, UCL-MSSL, UKATC, Univ. Sussex (UK); and Caltech, JPL, NHSC, Univ. Colorado (USA). This development has been supported by national funding agencies: CSA (Canada); NAOC (China); CEA, CNES, CNRS (France); ASI (Italy); MCINN (Spain); SNSB (Sweden); STFC (UK); and NASA (USA).}

\bibliographystyle{apj}
\bibliography{references}

\end{document}